\documentclass{aastex631}
\usepackage{amsmath}
\usepackage{threeparttable} 
\begin{document}

\title{Energy Cascade and Damping in Fast-Mode Compressible Turbulence}

\author[0000-0001-7205-2449]{Chuanpeng Hou}
\affiliation{Institut für Physik und Astronomie, Universität Potsdam, D-14476 Potsdam, Germany}

\author[0000-0003-2560-8066]{Huirong Yan}
\affiliation{Deutsches Elektronen Synchrotron (DESY), Platanenallee 6, D-15738 Zeuthen, Germany}
\affiliation{Institut für Physik und Astronomie, Universität Potsdam, D-14476 Potsdam, Germany}

\author[0000-0003-4268-7763]{Siqi Zhao}
\affiliation{Deutsches Elektronen Synchrotron (DESY), Platanenallee 6, D-15738 Zeuthen, Germany}
\affiliation{Institut für Physik und Astronomie, Universität Potsdam, D-14476 Potsdam, Germany}

\author[0000-0003-3400-191X]{Parth Pavaskar}
\affiliation{Deutsches Elektronen Synchrotron (DESY), Platanenallee 6, D-15738 Zeuthen, Germany}
\affiliation{Institut für Physik und Astronomie, Universität Potsdam, D-14476 Potsdam, Germany}

\correspondingauthor{Huirong Yan}
\email{huirong.yan@desy.de}

\begin{abstract}
Compressible turbulence governs energy transfer across scales in space and astrophysical systems. Capturing both the turbulence cascade and damping is therefore crucial for models of energy conversion, plasma heating, and particle transport in diverse plasma environments, but remains challenging. Progress is constrained by two unresolved fundamental questions: the persistence of the turbulence cascade in the presence of shocks and discontinuities, and the validity of classical wave theories under strong nonlinearity. In particular, it remains unclear whether meaningful cascade dynamics can be defined in compressible turbulence with phase steepening, and whether frameworks developed for monochromatic waves remain applicable to complex, broadband fluctuations. Using large-scale, high-resolution kinetic simulations, we analyze turbulence–particle interactions, which are beyond the capability of standard magnetohydrodynamic (MHD) simulations. We show that compressible turbulence damping at MHD scales in quantitative agreement with transit-time damping theory, even in fully developed nonlinear states. Moreover, the cascade persists despite the generation of shocks and discontinuities due to phase steepening, revealing a surprising robustness of cross-scale energy transfer under extreme conditions. We further provide the spectral expression of compressible turbulence. These results close a long-standing gap in the physics of compressible turbulence and establish a robust foundation for turbulence modeling from the heliosphere to galaxies.
\end{abstract}


\section{Introduction}
Turbulence is a fundamental phenomenon in space and astrophysical plasmas, playing a crucial role in many physical processes from heating and particle acceleration, transport of particles and energy, magnetic dynamo and reconnection, star formation and disk accretion, interstellar dust and chemistry, etc. \citep[][and references therein]{mac2004control,scalo2004interstellar, brandenburg2005astrophysical,retino2007situ, YLP08,Lazarian2009,Hirashita:2010ve, Ge16, comisso2018particle, ZhangYan2011, maiti2022cosmic,Yan2022, Zhao2025}. At magnetohydrodynamic (MHD) scales, turbulence is primarily governed by three low frequency propagating eigenmodes \citep{stix1992waves}, Alfv\'en-modes, fast-modes, and slow-modes, exhibiting distinct physical properties. Among them, incompressible Alfv\'en-mode turbulence, showing anisotropic cascade behavior perpendicular and parallel to background magnetic field with Kolmogorov scaling applicable only in the perpendicular direction, has been extensively studied \citep{goldreich1995toward,tu1995mhd,bruno2013solar,franci2018solar,yang2023energy}, and slow-mode turbulence is compressible but shares similarities with Alfv\'enic turbulence \citep{goldreich1995toward, lithwick2001compressible, cho2002compressible, makwana2020properties}. In contrast, compressible fast-mode turbulence remains the least understood. Existing studies suggest an almost isotropic cascade with a power-law index near $-1.5$ \citep{cho2002compressible, chandran2005weak, zhao2022multispacecraft, galtier2023fast}, but its dynamical robustness and damping process are still debated.

This knowledge gap arises for two main reasons. First, in-situ solar wind measurements usually show that Alfv\'en mode dominates the fluctuations, and numerous studies have investigated their properties and effects \citep{hou2024connecting,hou2024origin,rivera2024situ,yang2025natural,hou2025fine}, while compressible components contribute only a small fraction \citep{zhu2020wave}. Second, compressibility introduces significant theoretical complexity, making it challenging to characterize spectral distributions and energy cascade processes. Nevertheless, turbulence in magnetized plasma is intrinsically compressible with finite fraction of magnetosonic modes, without which a complete picture of MHD turbulence cascade can not be established \citep{cho2003compressible,makwana2020properties,Zhao2021}. Notably, fast-mode turbulence exerts the most significant impact on energetic particle transport and cosmic ray scattering \citep{yan2002scattering,yan2008cosmic}, making its dynamics central to astrophysical modeling.

The solar wind serves as a natural turbulence laboratory has demonstrated the importance of achieving a comprehensive understanding of compressible turbulence cascades. This importance is reflected in the application of the classical Yaglom law \citep{yaglom1949local}, which was extended to the solar wind to estimate the inertial energy cascade rate. To incorporate density fluctuations, a compressible extension based on density-weighted Elsässer variables was proposed, revealing a substantial amplification of the cascade rate \citep{carbone2009scaling}. On a firmer theoretical foundation, an exact law for compressible, isothermal MHD turbulence was subsequently derived \citep{banerjee2013exact} and successfully applied to THEMIS/ARTEMIS measurements, demonstrating that compressibility significantly modifies the cascade, particularly in the slow solar wind \citep{hadid2017energy}.

Beyond the cascade itself, its survival depends on the damping of compressible fluctuations. Compressibility introduces distinctive multiscale damping pathways. Magnetic mirror structures enable wave–particle interactions such as transit-time damping (TTD), which can act even at MHD scales and reshape the turbulence spectrum \citep{yan2002scattering,yan2004cosmic,petrosian2006damping,zhao2024small}. Meanwhile, longitudinal fast-mode fluctuations are susceptible to nonlinear steepening, producing shocks that may disrupt or redirect the cascade \citep{suzuki2006collisionless}. This inherent duality—between damping and cascade survival—has fueled long-standing debate over \textit{whether compressible turbulence can maintain a robust cascade under strong nonlinear conditions}, and \textit{whether classical theories such as TTD remain valid when confronted with intermittent, broadband fluctuations} \citep{cho2002compressible,yan2004cosmic,suzuki2006collisionless,suzuki2007cascading}.

These unresolved issues are not confined to a special case but are universal across plasma environments. This balance between cascade and damping dictates how turbulent energy is redistributed across scales and thereby controls the dynamics and observable properties of diverse systems. In the heliosphere, the competition between cascade and damping governs the transfer of energy from large MHD scales to kinetic scales \citep{bieber1996dominant,horbury2005spacecraft,osman2009quantitative,bruno2013solar,zank2020spectral}, regulating plasma heating, particle acceleration, and the variability of the solar wind that drives space weather. In the larger scale interstellar medium, fast-mode turbulence sets the scattering efficiency of high-energy cosmic rays, with cascade and damping jointly determining their diffusion coefficients, confinement times, and galactic spectra \citep{yan2002scattering,yan2008cosmic}. Across all these systems, the absence of a clear description of the cascade–damping balance limits our ability to model plasma turbulence consistently from the heliosphere to galaxies.

Addressing these issues require methods that can overcome the intrinsic limitations of current observational and numerical tools. Specifically, spacecraft measurements, limited to single points or sparse cluster, cannot accurately separate the temporal and spatial evolution of turbulence. Telescopes, in turn, offer only line-of-sight integrated data, which precludes a global reconstruction. Meanwhile, conventional MHD simulations can capture the overall dynamics of turbulence but neglect particle feedback on electromagnetic fields and thus fail to assess damping in the inertial range.

In this work, we perform large-domain, high-resolution hybrid and fully kinetic particle-in-cell (PIC) simulations designed to investigate fast-mode dominated compressible turbulence. These simulations allow a fully self-consistent investigation showing that MHD-scale fluctuations dissipate in agreement with TTD theory and that energy cascade persists despite strong wave steepening. By resolving these long-standing uncertainties, our findings provide decisive insight into the fundamental physics of compressible turbulence and establish a robust foundation for future theoretical and observational studies.

\section{Simulation Methodology}\label{simulation_setup}

We perform numerical simulations of compressible turbulence dominated by fast magnetosonic modes using two complementary simulation frameworks: hybrid  \citep{bowers2008ultrahigh,le2021astrophysical} and PIC \citep{bowers20080,bowers2008ultrahigh,bowers2009advances} simulations. In both cases, the initial fluctuations of electric fields, magnetic fields, and particle moments are constructed based on the polarization of fast-modes. The systems then evolve without external forcing, resulting in a decaying turbulence process. 
The primary distinction between the two simulation frameworks lies in the treatment of electrons. In the hybrid model, only protons are treated as kinetic particles, while electrons are modeled as a massless fluid. In contrast, the fully kinetic PIC simulation treats both protons and electrons as kinetic particles, with a reduced proton-to-electron mass ratio ($m_\mathrm{p}/m_\mathrm{e}=20$). This difference in electron treatment (fluid vs. kinetic) effectively isolates different damping processes in the fast modes turbulence. 

To capture turbulent structures at MHD scales, both simulations use large computational domains with a periodic boundary conditions, see Table \ref{tab:comparison} for a summary of the following simulation parameters. The three-dimensional (3D) hybrid simulation domain has a size of $1024\,d_\mathrm{i} \times 1024\,d_\mathrm{i} \times 1024\,d_\mathrm{i}$, where $d_\mathrm{i} = v_\mathrm{A}/\Omega_\mathrm{p}$  is the proton inertial length, $v_\mathrm{A}$ is the Alfv\'en speed, and $\Omega_\mathrm{p}$ is the proton gyrofrequency. The number of grid cells is $512^3$, corresponding to a spatial resolution of $2\,d_\mathrm{i}/grid$. Each grid cell contains 500 proton particles, yielding a total of $512^3 \times 500$ particles. The simulation time step is $\Delta t = \,0.1\,\Omega_\mathrm{p}^{-1}$. For the two-dimensional (2D) fully kinetic PIC simulations, due to computational resource limits, the domain size is $1024\,d_\mathrm{i} \times 1024\,d_\mathrm{i}$, with $1024^2$ grid cells (i.e., $1\,d_\mathrm{i}/grid$ resolution). Each cell contains 500 proton and 500 electron particles, resulting in a total particle number of $2 \times 1024^2 \times 500$. The time step is set to $\Delta t = \,0.09\,\Omega_\mathrm{e}^{-1}$, where $\Omega_\mathrm{e}$ is the electron gyrofrequency. 
Small time steps ensure accurate tracking of particle motion, especially that of electrons in the PIC simulation. We also conducted a higher-resolution PIC simulation with a domain size of $512d_\mathrm{i} \times 512 d_\mathrm{i}$, resolved by $4096^2$ grid cells, yielding a spatial resolution of $0.125 d_\mathrm{i}/grid$. For all types of simulations, the particle velocity consists of three orthogonal components.

Turbulence is initialized by superposing fast-modes whose wave-vectors along each simulation dimension ($k_\mathrm{0},2k_\mathrm{0},3k_\mathrm{0}$) are integer multiples of the base wavenumber $k_\mathrm{0} = 2\pi / L_\mathrm{0}$, where $L_\mathrm{0}$ is the simulation box size. The integer wave-vectors avoid artificial field and density discontinuities at the domain boundaries. The electric and magnetic field fluctuations, as well as particle moments, are initialized with random phases and equal amplitudes for all wave-vectors, producing a domain-averaged fluctuation amplitude $\delta v_{\mathrm{rms}} / v_\mathrm{A} \approx 0.23$, where $\delta v_{\mathrm{rms}} = \sqrt{\left<|\delta v|^2\right>}$ denotes the root-mean-square of velocity fluctuation $\delta v$.

In the 3D hybrid simulations, the background magnetic field is aligned along the $z$-axis ($\mathbf{B}_\mathrm{0} = B_\mathrm{0}\,\hat{\mathbf{z}}$), and the fast-mode waves are initialized in 3D $\mathbf{k}$-space. In the 2D PIC simulations, the background field is along the $x$-axis ($\mathbf{B}_\mathrm{0} = B_\mathrm{0}\,\hat{\mathbf{x}}$), and waves are only injected in the x-y plane. For all simulations, the plasma $\beta$, defined as the ratio of thermal to magnetic pressure, is set to $\beta = 0.2$, and the initial proton and electron temperatures are equal, i.e., $T_\mathrm{p} = T_\mathrm{e}$. All simulations were performed on the NHR@ZIB high-performance computing cluster.

At MHD scales, both numerical simulations and in-situ spacecraft measurements have revealed the presence of mode conversion among different eigenmodes \citep{cho2003compressible,grant2018NatPhy,reville2018ApJ}. Although our simulations are initialized only with fast modes, the subsequent nonlinear evolution inevitably generates Alfv\'enic and slow-mode components, albeit with lower energy compared to the dominant fast-mode fluctuations. Given that the primary objective of this study is to characterize the evolution of fast-mode dominated compressible turbulence, it is essential to effectively distinguish the contributions from fast-modes within the turbulence. To achieve this, we adopt the classical method introduced by \citep{cho2002compressible,zhao2024small}.
The wave-mode decomposition exploits the mutual orthogonality of the eigenmode velocity vectors—corresponding to Alfv\'en, slow, and fast modes—in the wavevector–magnetic field coordinate system. By projecting the turbulent velocity fluctuations onto the respective displacement vectors (Appendix \ref{appendix_decomp}), we are able to quantitatively extract the fast-mode component from the mixed turbulent field.

\begin{table}[ht]
\caption{\label{tab:comparison}
The setup for 3D Hybrid and 2D Kinetic PIC Simulations.}
\begin{ruledtabular}
\begin{tabular}{cccc}
& 3D Hybrid  & 2D PIC  & 2D  PIC(high resolution) \\
\hline
Simulation Code & Hybrid-VPIC & VPIC  & VPIC\\
Electron Treatment & Fluid (massless) & Kinetic ($m_\mathrm{p}/m_\mathrm{e} = 20$) & Kinetic ($m_\mathrm{p}/m_\mathrm{e} = 20$)\\
Spatial Dimension & 3D (x, y, z) & 2D (x, y) & 2D (x, y)\\
Velocity Space Dimension & 3D ($V_x$, $V_y$, $V_z$) & 3D ($V_x$, $V_y$, $V_z$) & 3D ($V_x$, $V_y$, $V_z$)\\
Domain Size & $1024\,d_\mathrm{i} \times 1024\,d_\mathrm{i} \times 1024\,d_\mathrm{i}$ & $1024\,d_\mathrm{i} \times 1024\,d_\mathrm{i}$ & $512\,d_\mathrm{i} \times 512\,d_\mathrm{i}$\\
Grid Resolution & $512^3$, i.e., $2\,d_\mathrm{i}/\text{grid}$ & $1024^2$, i.e., $1\,d_\mathrm{i}/\text{grid}$ & $4096^2$, i.e., $0.125\,d_\mathrm{i}/\text{grid}$\\
Particles per Cell & 500 protons & 500 protons + 500 electrons & 500 protons + 500 electrons\\
Total Particles & $512^3 \times 500$ & $2 \times 1024^2 \times 500$ & $2 \times 4096^2 \times 500$\\
Time Step & $\Delta t = 0.1\,\Omega_\mathrm{p}^{-1}$ & $\Delta t = 0.09\,\Omega_\mathrm{e}^{-1}$ & $\Delta t = 0.09\,\Omega_\mathrm{e}^{-1}$\\
Boundary Condition & Periodic & Periodic & Periodic\\
Magnetic Field Direction & $\mathbf{B}_0 = B_0\,\hat{\mathbf{z}}$ & $\mathbf{B}_0 = B_0\,\hat{\mathbf{x}}$ & $\mathbf{B}_0 = B_0\,\hat{\mathbf{x}}$\\
Initial Fluctuations & Fast-mode in 3D $\mathbf{k}$-space & Fast-mode in 2D $\mathbf{k}$-space& Fast-mode in 2D $\mathbf{k}$-space\\
Driving & Decaying turbulence  & Decaying turbulence & Decaying turbulence\\
Plasma Beta & $\beta = 0.2$ & $\beta = 0.2$ & $\beta = 0.2$\\
Initial Temperature & $T_\mathrm{p} = T_\mathrm{e}$ & $T_\mathrm{p} = T_\mathrm{e}$ & $T_\mathrm{p} = T_\mathrm{e}$ \\
Fluctuation Amplitude & $\delta v_{\mathrm{rms}} / v_\mathrm{A} \approx 0.23$ & $\delta v_{\mathrm{rms}} / v_\mathrm{A} \approx 0.23$  & $\delta v_{\mathrm{rms}} / v_\mathrm{A} \approx 0.21$\\
\end{tabular}
\end{ruledtabular}
\end{table}

\section{Results and Discussion}

\subsection{Cascade physics of compressible turbulence}

\begin{figure}[ht]
\centering
\includegraphics[width=1.0\textwidth]{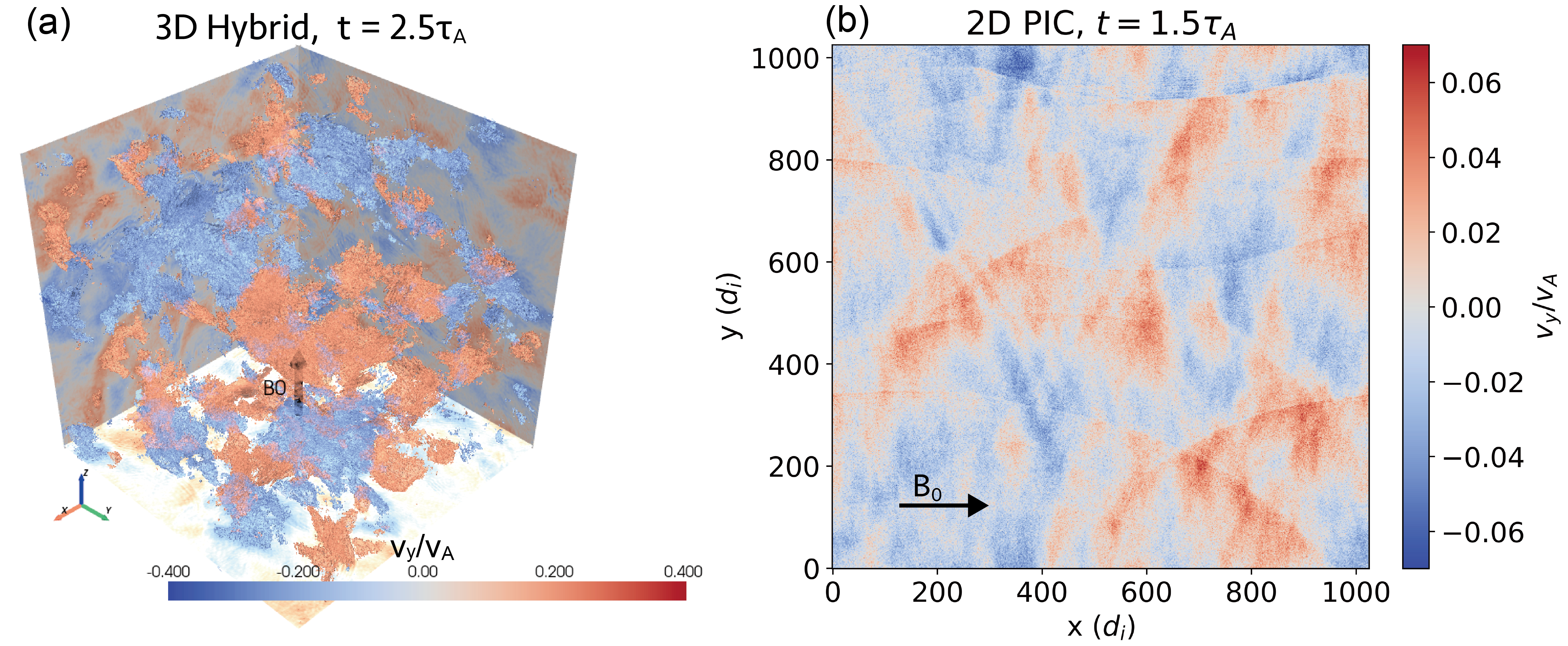}
\caption{Spatial distribution of near-isotropic features in the compressible turbulence. (a) Isosurfaces of the velocity component $v_\mathrm{y}$ from the 3D hybrid simulation, plotted at levels -0.2$v_A$ and 0.2$v_A$. The color bar indicates the range of $v_\mathrm{y}$ values across three orthogonal slices. (b) Distribution of proton $v_\mathrm{y}$ in the 2D PIC simulation. The direction of background magnetic field is shown by black arrows.}\label{fig1} 
\end{figure}

Figure \ref{fig1} presents the isosurfaces of velocity fluctuations ($v_y$) from the three-dimensional (3D) hybrid simulation at $ t = 2.5\,\tau_{\mathrm{A}} $, where the Alfv\'en transit time is defined as $\tau_\mathrm{A} = L_\mathrm{0}/v_\mathrm{A}$, 
with $ L_{0} $ the simulation domain size and $ v_{\mathrm{A}} $ the Alfv\'en speed. For comparison, $v_y$ fluctuations from the two-dimensional (2D) fully kinetic PIC simulation are shown at $ t = 1.5\,\tau_{\mathrm{A}} $. In both panels, the direction of the background magnetic field $ \mathbf{B}_{0} $ is indicated. 
Remarkably, despite the differences in dimensionality and kinetic treatment of electrons, both simulations exhibit highly consistent turbulent morphologies. Plasma eddies appear as nearly isotropic in contrast to the filamentary and strongly anisotropic patterns typically associated with Alfv\'enic turbulence shown in spacecraft measurements \citep[see][]{horbury2008anisotropic,chen2012three,verdini20183d}, MHD simulations \citep[e.g.,][]{lazarian2024sub} and kinetic simulation \citep[e.g.,][]{franci2018solar}. It is important to note that the velocity fields shown in Figure \ref{fig1} have not undergone wave-mode decomposition.  The near-isotropy is a characteristic signature of fast-mode dominated turbulence, reflecting its inherently weak anisotropy in the cascade process. This visual agreement between hybrid and PIC simulations directly confirms that fast modes dominate the turbulent fluctuations in our setups.

\begin{figure}[ht]
\centering
\includegraphics[width=1.0\textwidth]{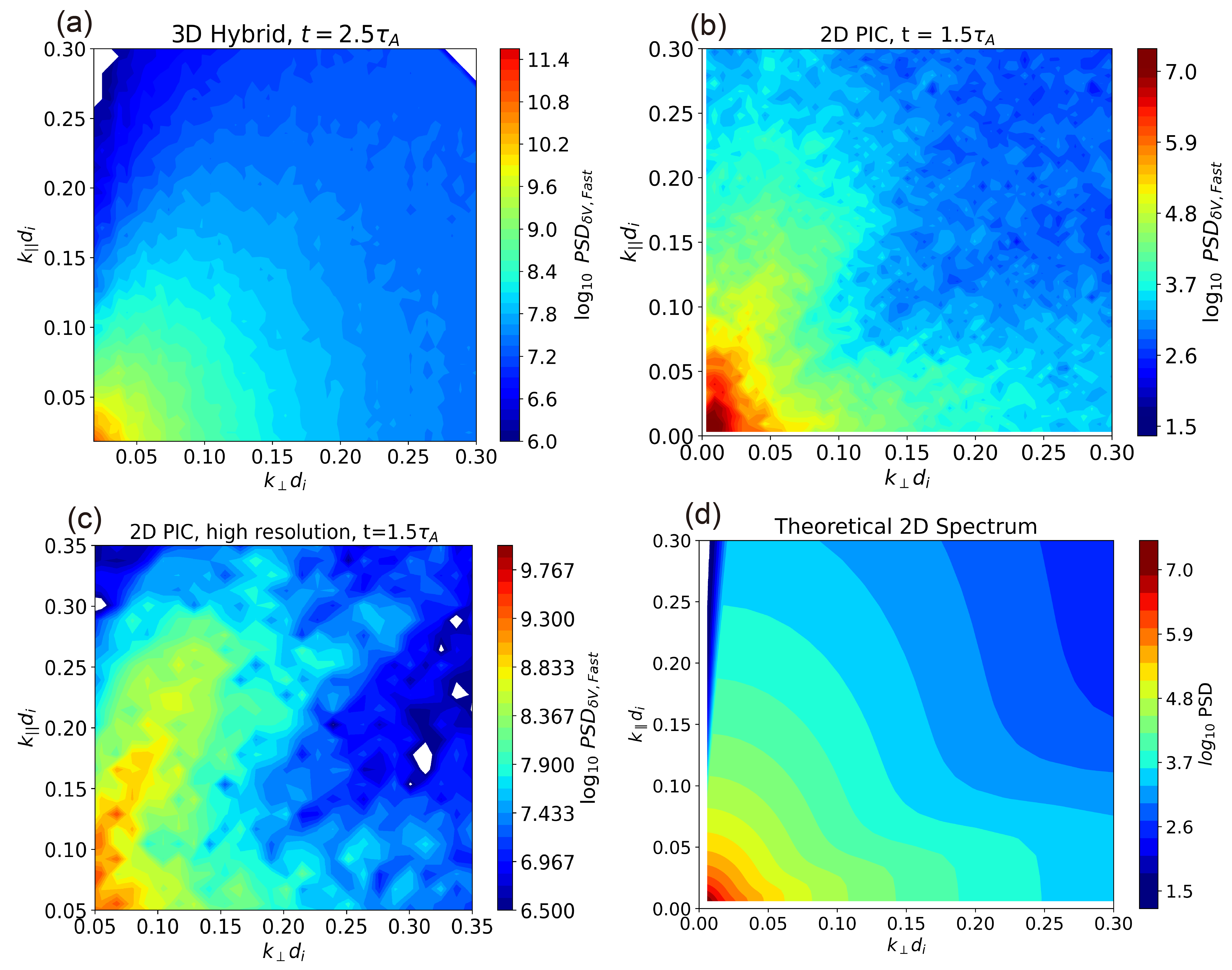}
\caption{Angle-dependent collisionless damping in the near-isotropic cascade of fast mode turbulence. (a) PSD distribution in $k_{\perp}-k_{\parallel}$ space  calculated from the 3D hybrid simulation results. (b) PSD distribution in $k_{\perp}-k_{\parallel}$ space calculated from the 2D PIC simulation. (c) PSD distribution in $k_{\perp}-k_{\parallel}$ space calculated from 2D PIC simulation with a higher resolution. (d) Theoretical PSD distribution in $k_{\perp}-k_{\parallel}$ space according to Equation \ref{eq_theory}.}\label{fig2} 
\end{figure}

To quantitatively isolate the fast-mode component from the mixed turbulent field, we first perform a Fourier transform of the velocity field, and then apply wave-mode decomposition in Fourier space based on the eigenmode displacement vectors of the linear MHD modes. This procedure separates the contributions from Alfv\'en, slow, and fast magnetosonic modes for each wavevector $\mathbf{k}$.
We then compute the power spectral density (PSD) of the extracted fast-mode fluctuations in the two-dimensional wavevector space, resolved into directions parallel $(\hat{\mathbf{k}}_\parallel)$ and perpendicular $(\hat{\mathbf{k}}_\perp)$ to the global background magnetic field $\mathbf{B}_0$ (Figures \ref{fig2}(a)-\ref{fig2}(c)). The hybrid simulation produces an approximately isotropic PSD, with only weak anisotropy arising from the reduced compressibility of quasi-parallel fast modes. In contrast, the PIC simulation exhibits a pronounced anisotropy: the PSD extends along both $\hat{\mathbf{k}}_\parallel$ and $\hat{\mathbf{k}}_\perp$, but displays a clear depression in power near a propagation angle $\theta \approx 55^{\circ}$, where $\theta$ is defined as the angle between the wavevector $\mathbf{k}$ and $\mathbf{B}_0$. This angular power depression is later shown to be associated with the TTD theory. 2D hybrid simulations are also performed as a comparison test. The results show isotropic PSD similar with Figure \ref{fig2}(a), confirming that the difference in the 3D hybrid and 2D PIC simulations is not due to the dimensionality. In other words, 2D PIC simulations capture the cascade physics owing to the fact that the fast modes cascade is primarily in the radial direction. The theoretical PSD (Figure \ref{fig2}(d)), showing angle-dependent anisotropy, is presented based on the Equation \ref{eq_theory} introduced in section \ref{sec_truncation}.

\begin{figure}[ht]
\centering
\includegraphics[width=1.0\textwidth]{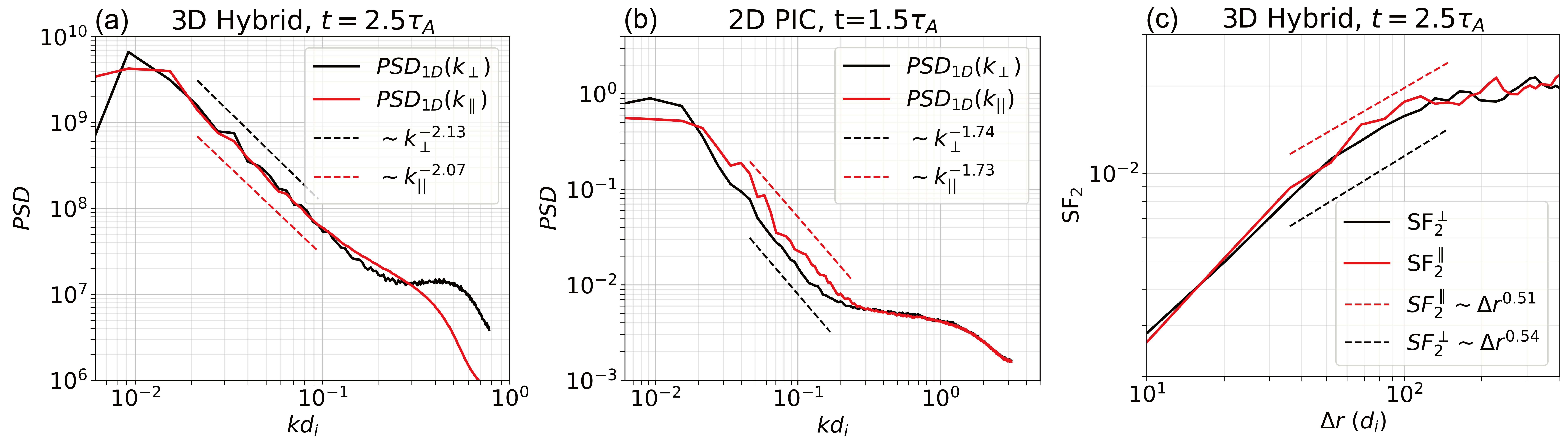}
\caption{The 1D energy spectrum of compressible turbulence.
(a) Integrated 1D PSD of the fast-mode velocity field as a function of $kd_{\mathrm{i}}$ obtained from the 3D hybrid simulation.
(b) Integrated 1D PSD of the fast-mode velocity field as a function of $kd_{\mathrm{i}}$ obtained from the 2D PIC simulation.
(c) Integrated 1D PSD of the fast-mode velocity field as a function of $kd_{\mathrm{i}}$ obtained from the high resolution 2D PIC simulation.
(d) Filtered $SF_{2,v}$ of velocity field from the 3D hybrid simulation, as defined by Equation \ref{filter_SF}. 
(e) $SF_{2,v}$ of velocity field from the 2D PIC simulation.
(f) $SF_{2,v}$ of velocity field from the high resolution 2D PIC simulation.} 
In all panels, the fitted scalings are shown as dashed lines.\label{fig3} 
\end{figure}

A key diagnostic of turbulence cascade behavior is the spectral index of the power spectral density (PSD). By integrating the two-dimensional PSD $P(k_\parallel, k_\perp)$ shown in Figures \ref{fig2}(a)-\ref{fig2}(c) over the perpendicular and parallel directions, respectively, we obtain the one-dimensional PSD (1D-PSD) as a function of the scalar wavenumber $k$. As shown in Figure \ref{fig3}(a), the hybrid simulation yields spectral indices close to $-2$ in both the parallel and perpendicular directions.  
The PIC simulation (Figures \ref{fig3}(b) and \ref{fig3}(c)) produces spectra with indices near $-1.7$.  
In both cases, the spectral slopes are steeper than the theoretical prediction of $-1.5$ for an undamped fast-mode cascade \citep{cho2002compressible,galtier2023fast}.

This steepening can be attributed to phase steepening, a nonlinear process that naturally occurs in compressible fast-mode fluctuations due to the nonzero projection $\mathbf{k} \cdot \delta \mathbf{v} \neq 0$. Here, the velocity perturbation $\delta \mathbf{v}$ has a component along $\mathbf{k}$, allowing nonlinear self-compression of the wave phases.  
This effect generates localized sharp gradients or shocks in physical quantities such as velocity and magnetic field. When such discontinuities are Fourier transformed, they steepen the measured spectral index in k space. To mitigate the influence of wave phase steepening on the measured spectral index, we replace the Fourier-based analysis with the second-order structure function \citep{wu2020isotropic} of the velocity field, defined as  
\begin{equation}
SF_2(\Delta \mathbf{r}) = \left< \left| \Delta v \right|^2 \right>_\mathbf{r},
\end{equation}
where $\left<... \right>_\mathbf{r}$ denotes the spatial average, $\Delta v = v(\mathbf{r} + \Delta \mathbf{r}) - v(\mathbf{r})$,
$\mathbf{r}$ is the position vector, and $\Delta \mathbf{r}$ is the spatial separation vector.  

For a power-law spectrum $PSD(k) \propto k^{-\alpha}$, the structure function scaling follows $SF_2 \propto \Delta r^{\gamma}$, with the exponents related by $\alpha = 1 + \gamma$.
For example, a PSD spectral index of $-1.5$ corresponds to $\alpha = 1.5$ and thus $\gamma = 0.5$.
The advantage of the structure function approach is that large-amplitude discontinuities, such as shocks or strong phase jumps, can be effectively removed using threshold-based filters applied to $|\Delta v|^2$. By suppressing the influence of such intermittent features, the method yields a more reliable estimate of the underlying inertial-range scaling that is directly associated with the turbulent cascade, rather than with localized steepening events.

Figure \ref{fig3}(d) presents the second-order structure functions after applying a jump-removal filter, defined as  
\begin{equation}\label{filter_SF}
SF_2 = \left< |\Delta v|^2_{\Delta v^2 < 0.1 \,\mathrm{max}(\Delta v^2)} \right>_\mathbf{r},
\end{equation}
which removes contributions from large-amplitude discontinuities such as shocks and sharp phase jumps.  
In both the parallel and perpendicular directions, the filtered $SF_2$ exhibits a scaling index of $\gamma \approx 0.5$, corresponding to a power spectral index of $-1.5$.  
This scaling is in excellent agreement with theoretical predictions for fast-mode turbulence \citep{cho2002compressible, galtier2023fast} and matches the expected cascade rate $\propto k^{0.5}$. The $SF_2$ analysis of 2D PIC simulation yields a slope of approximately 0.3 (Figure \ref{fig3}(e)), which is flatter than the slope derived from the Fourier-based PSD. This discrepancy stems from that PIC simulations must resolve small-scale electron dynamics and lack the smoothing inherent in electron fluid approximations, resulting in higher intrinsic noise levels. The presence of small-scale noise increases the values of the $SF_2$ at small separations, thereby flattening the measured spectrum. By contrast, the $SF_2$ from higher-resolution PIC simulations appears consistent with expectations (Figure \ref{fig3}(f)), as the reduced noise produces a more reliable slope.

In general, the structure function analysis confirms that the intrinsic inertial-range scaling of the cascade remains valid, demonstrating that phase steepening does not terminate the energy cascade process.
A similar cascade behavior is expected in realistic environments such as the solar wind, where turbulence is dominated by Alfv\'en modes, with strong turbulence being a key observational feature often interpreted in terms of critical balance\citep{goldreich1995toward,horbury2008anisotropic}. Within this framework, weak wave turbulence theory is still applicable to fast modes, owing to their nearly isotropic propagation, higher phase speed, and longer nonlinear interaction time compared to Alfvén modes.

\subsection{Collisionless damping}

The power spectral dip in Figures \ref{fig2}(b) and \ref{fig2}(c) aligns well with linear theoretical predictions:  
$\theta \approx 55^{\circ}$ corresponds to the characteristic propagation angle at which fast magnetosonic waves experience their strongest resonant damping via TTD.  
At this angle, the wave phase speed $ v_{\mathrm{ph},\parallel} = v_{\mathrm{ph}} \cos\theta $ matches the thermal velocity of electrons, maximizing the efficiency of wave--particle interactions.
From linear theory, the TTD damping rate \citep{ginzburg1962propagation,yan2004cosmic} in a low-$\beta$ plasma is given by  
\begin{equation}\label{eq_damping}
\Gamma = k v_{\mathrm{ph}} \frac{\sqrt{\pi \beta}}{4} \frac{\sin^2 \theta}{\cos \theta}
        \left[ \sqrt{\frac{m_\mathrm{e}}{m_\mathrm{p}}} 
        \exp \left( -\frac{m_\mathrm{e}/m_\mathrm{p}}{\beta \cos^2 \theta} \right)  + 5 \exp \left( -\frac{1}{\beta \cos^2 \theta} \right) \right],
\end{equation}
where $ v_{\mathrm{ph}} $ is the fast-mode phase speed, $ m_\mathrm{p} $ is the proton mass, and $ m_\mathrm{e} $ is the electron mass.  
The first exponential term represents damping dominated by resonant electrons, while the second term corresponds to damping caused by resonant protons.

For the reduced mass ratio $ m_\mathrm{p} / m_\mathrm{e} = 20 $ used in our PIC simulation, Equation \ref{eq_damping} predicts a maximum damping rate precisely at $\theta \approx 55^{\circ}$, in quantitative agreement with the angular position of the PSD depression observed in Figures \ref{fig2}(b) and \ref{fig2}(c). This consistency confirms that TTD plays a dominant role in shaping the angular distribution of power in fast-mode turbulence.
On the other hand, no such TTD-related damping features are observed in the hybrid simulation (Figure \ref{fig2}(a)).  
The underlying reason lies in the resonance condition for TTD, which requires that particles have parallel velocities close to the wave’s parallel phase speed, $v \approx v_{\mathrm{ph}}$.
In the hybrid simulation, only proton kinetics are included.  
For the low-$\beta$ ($\beta = 0.2$) plasma considered here, the proton thermal speed is  
$v_{\mathrm{th},p} \approx 0.22\,v_{\mathrm{ph}}$, 
which is significantly lower than the fast-mode phase speed $ v_{\mathrm{ph}} $.  
As a result, the fraction of protons that satisfy the resonance condition is extremely small, leading to negligible TTD damping.  
In contrast, in the PIC simulation, electrons with lower mass have a higher thermal speed,  
$v_{\mathrm{th},e} \approx v_{\mathrm{ph}}$, 
allowing a substantial portion of the electron population to fulfill the resonance condition.  
Consequently, MHD-scale turbulent damping mediated by TTD manifests prominently only in the PIC simulations, where electron kinetics are fully resolved.

\subsection{Cascade truncation}\label{sec_truncation}

We identify the truncation scale directly from the PIC simulations. In the absence of TTD, the two-dimensional power spectral density (2D-PSD) of fast modes is expected to be nearly isotropic, reflecting the weak intrinsic anisotropy of the fast-mode cascade. This behavior is indeed observed along quasi-parallel and quasi-perpendicular propagation directions, where the TTD damping rate is minimal and the inertial range extends down to $ k d_i \approx 0.3 $ (Figures \ref{fig2}(b) and \ref{fig2}(c)). In contrast, at the propagation angle where TTD damping is predicted to be strongest ($ \theta \approx 55^\circ $), resonant interactions efficiently extract energy from the cascade, causing the inertial range to terminate much earlier, with a spectral break already appearing at $ k d_i \approx 0.12 $ (Figures \ref{fig2}(b) and \ref{fig2}(c)). This pronounced angular dependence---absent in the undamped cascade---indicates that the break scale is not a natural outcome of nonlinear energy transfer, but rather a consequence of enhanced collisionless damping. We therefore interpret this anisotropic cutoff as the TTD-induced truncation scale, which directly links the energy-containing range of the turbulence to the resonance condition between the fast-mode phase speed $ v_{\mathrm{ph}} $ and particle parallel velocities $ v_\parallel $. Such a link implies that the damping scale, and thus the partition of turbulent energy into particle heating, can be predicted from the local plasma parameters that control $ \Gamma $.

To validate the angular-dependent spectral damping observed in the PIC simulations, we derive a theoretical expression for the turbulence spectrum.  
Starting from the cascade rate for isotropic fast-mode turbulence \citep{yan2004cosmic},  $
\tau_{\mathrm{cas}}^{-1} \approx \left( k/L \right)^{1/2} \delta v^2/v_{\mathrm{ph}}$, 
and incorporating the weak angular anisotropy of the cascade along different propagation angles \citep{galtier2023fast}, we modify the expression to  
\begin{equation}\label{eq_cascade}
\tau_{\mathrm{cas}}^{-1} \approx C \left( \frac{1 + 2\sin^2\theta}{\sin\theta} \right)
\left( \frac{k}{L} \right)^{1/2} \frac{\delta v^2}{v_{\mathrm{ph}}},
\end{equation}
where $C$ is a dimensionless constant to be determined from the simulations, L is the minimum injection scale.

By incorporating the TTD damping rate $\Gamma(k,\theta)$ into the steady-state turbulence energy balance equation \citep{petrosian2006damping}, we obtain the angular-dependent turbulence spectrum (see Appendix \ref{appendix_theory} for details):  
\begin{equation}\label{eq_theory}
PSD(k,\theta) \approx k^{-5/2} \exp\left( -\frac{\Gamma(k,\theta)}{\tau_{\mathrm{cas}}^{-1}(k,\theta)} \right).
\end{equation}

The resulting theoretical angular-dependent spectrum is shown in Figure \ref{fig2}(d) and exhibits agreement with the PIC results in Figures \ref{fig2}(b) and \ref{fig2}(c).  
According to Equation \ref{eq_theory}, the cascade truncation scale $k_{\mathrm{tr}}(\theta)$ can be determined by equating the cascade rate with the damping rate,  
\begin{equation}
\tau_{\mathrm{cas}}^{-1}(k_{\mathrm{tr}},\theta) = \Gamma(k_{\mathrm{tr}},\theta).
\end{equation}
Combining this relation with the truncation scales measured from our 2D PIC simulations (Figures \ref{fig2}(b) and \ref{fig2}(c)), we determine $C \approx 1.1$ in Equation \ref{eq_cascade}.

As shown in Figure \ref{fig3}(b), the PSD flattens when $kd_i > 0.2$. This phenomenon can be attributed to the limited ability of low–spatial-resolution simulations to resolve ion-scale dissipation processes and stronger thermal noise. The flattened region likely obscures the truncation scale of the energy cascade. To address this, we performed higher–spatial-resolution simulations (see Table \ref{tab:comparison}), increasing the grid resolution to $0.125d_i$. The high-resolution results are broadly consistent with those of the low-resolution PIC simulation. In the one-dimensional PSD, the truncated features of the energy cascade are clearly visible, as the spectral slope steepens from $k^{-1.7}$ to $k^{-2.45}$ at $kd_i \approx 0.12$.

Under the realistic mass ratio ($ m_\mathrm{p} / m_\mathrm{e} = 1836 $), quasi-perpendicularly propagating fast modes experience the strongest transit-time damping, with the peak damping rate occurring at a propagation angle of $ \theta \approx 86^{\circ} $. In the 2D-PSD, this manifests as a pronounced suppression of power along the quasi-perpendicular direction, while quasi-parallel regions retain their power across a broad inertial range. The corresponding 1D-PSD from Equation \ref{eq_theory} reveals that the parallel spectrum maintains a spectral index close to $-1.5$, consistent with theoretical predictions for an undamped cascade. In contrast, the perpendicular spectrum slightly steepens. Our study provide direct evidence for the dominant role played by collisionless damping in truncating the cascade of fast modes, and angle-dependent modulation of compressible turbulence.

\subsection{Particle heating}

\begin{figure}[ht]
\centering
\includegraphics[width=0.9\textwidth]{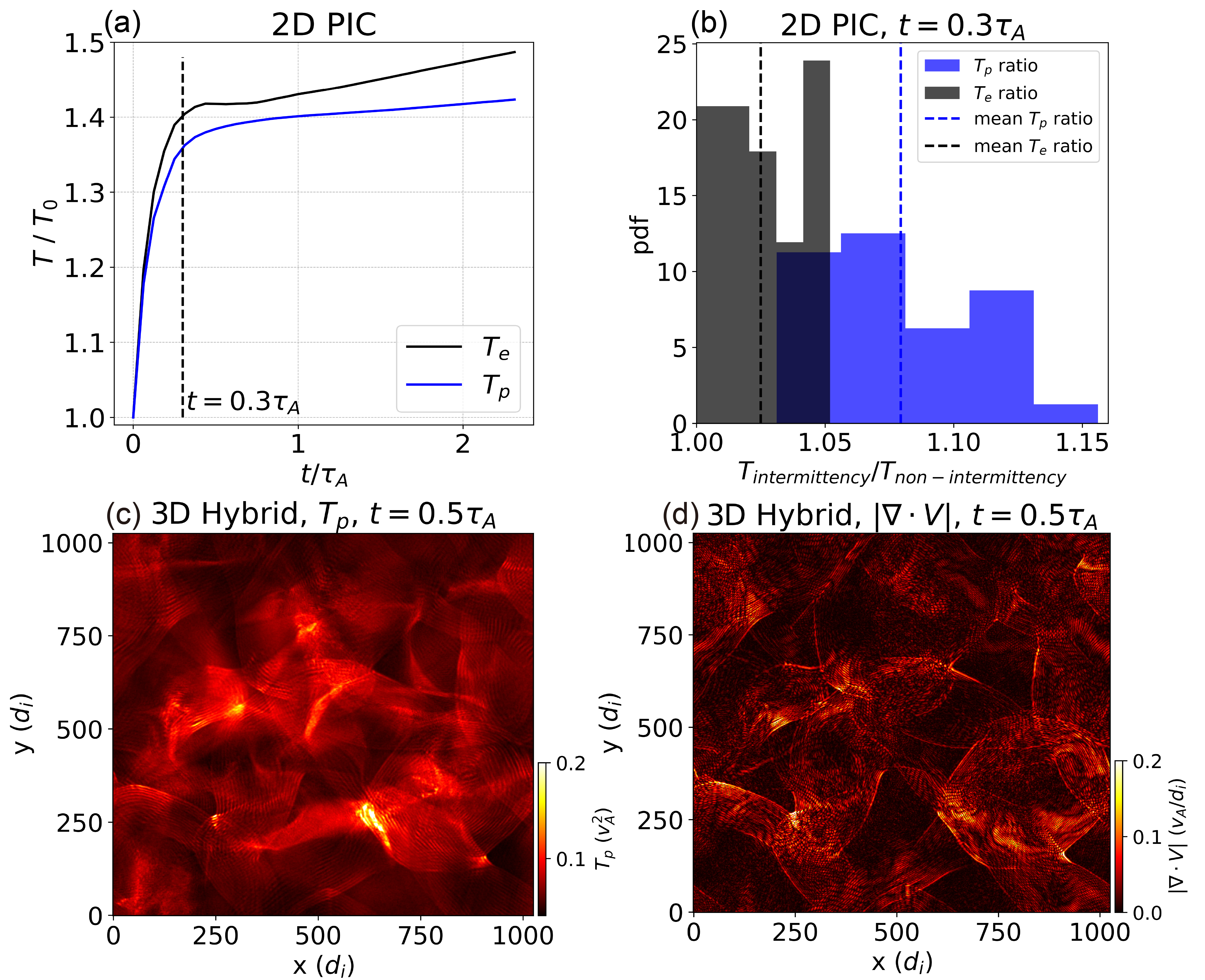}
\caption{Energy source for particle heating. (a) Temporal evolution of proton and electron temperatures from PIC simulation. The dashed line marks the time corresponding to panel (b). (b) Probability density function (pdf) of temperature ratio between intermittent and non-intermittent regions. The mean temperature ratio is shown by dashed lines. (c) Proton temperature in the perpendicular magnetic field ($x$–$y$) plane, sliced at $z = 0 ~d_\mathrm{i}$ from 3D hybrid simulation. 
(d) Same as (c) but for the divergence of velocity field.}\label{fig4} 
\end{figure}

Now we consider the energy source for particle heating. 
To distinguish between contributions from coherent intermittent structures and the background turbulent cascade, we perform a combined analysis of the second- and fourth-order structure functions for both the velocity and magnetic fields. Specifically, we compute the kurtosis,
\begin{equation}
K_v = \frac{\left< \Delta v^4 \right>}{\left< \Delta v^2 \right>^2}, \quad
K_B = \frac{\left< \Delta B^4 \right>}{\left< \Delta B^2 \right>^2},
\end{equation}
where $\Delta v = v(\mathbf{r} + \mathbf{\Delta r}) - v(\mathbf{r})$, 
$\Delta B = B(\mathbf{r} + \mathbf{\Delta r}) - B(\mathbf{r})$,
and $\mathbf{\Delta r}$ is the spatial separation vector. 
A kurtosis value $ K > 3 $ corresponds to a non-Gaussian distribution with heavy tails, which is a common signature of intermittent, jump-like structures such as shocks or current sheets. 

In the 2D PIC simulation, regions with $ K_v > 3 $ occupy less than $ 17\% $ of the total simulation volume, while regions with $ K_B > 3 $ occupy less than $ 15\% $. 
These relatively small volume fractions indicate that the dominant mechanism of cross-scale energy transfer is a continuous turbulent cascade, rather than sporadic dissipation in localized intermittent structures. 
This also suggests that particle heating in the simulation is primarily driven by the broadband cascade and associated resonant interactions, with intermittency providing only a secondary contribution.

 For the 2D PIC simulation, the electron temperature increases over time at a substantially faster rate than that of the protons (Figure \ref{fig4}(a)), reflecting the electrons’ ability to resonate efficiently with a broad range of turbulent fluctuations. A comparison of temperature evolution between intermittent and non-intermittent regions shows that protons undergo preferential heating within intermittent, jump-like structures, whereas the electron temperature exhibits no clear correlation with intermittency (Figure \ref{fig4}(b)). This difference arises because electron heating is dominated by resonant wave-particle interactions with broadband fluctuations generated by the turbulent cascade, rather than by localized dissipation events.

Similarly, in the 3D hybrid simulation, the low proton thermal velocity $ v_{\mathrm{th},p} $ relative to the fast-mode phase speed $ v_{\mathrm{ph}} $ suppresses efficient resonance, resulting in proton heating being almost concentrated near regions of strong velocity-field divergence (Figures \ref{fig4}(c) and \ref{fig4}(d)), consistent with compression-driven heating mechanisms. This contrast between electron and proton heating provides direct evidence for the coexistence of a large-scale turbulent cascade and TTD, even in steepened compressible turbulence, and highlights the distinct pathways through which different particle species gain energy in collisionless plasmas.

\section{Conclusion}

In this work, we provide the first self-consistent evidence that fast-mode–dominated compressible turbulence sustains a cross-scale cascade while dissipating in quantitative agreement with transit-time damping theory, even in the presence of strong nonlinearity and shock formation. Large-domain hybrid and fully kinetic PIC simulations clarify the distinct roles of proton and electron kinetics in shaping spectral distributions and truncation scales, and directly connect angle-dependent damping to collisionless resonances. In parallel, our derivation of the spectral expression for fast-mode turbulence establishes a direct theoretical counterpart to the simulations. These results form a unified numerical–theoretical framework that provides a solid basis for interpreting spacecraft observations, guiding laboratory experiments, and improving global models across a wide range of astrophysical environment. Beyond resolving long-standing debates on cascade persistence and damping mechanisms, our findings build a foundation for a deeper understanding of energy transfer, dissipation, and particle heating in collisionless plasmas.

\begin{acknowledgments}
We acknowledge the computing resources from the high-performance computers at the NHR center NHR@ZIB, with the project No. bbp00080.
The hybridVPIC simulation code is available at \url{https://github.com/lanl/vpic-kokkos/tree/hybridVPIC} and VPIC simulation code is available at \url{https://github.com/lanl/vpic}.
\end{acknowledgments}

\appendix

\section{Calculation of power spectrum density}

For 3D-hybrid simulation, we perform a three-dimensional Fourier transform of the velocity field to obtain its spectral representation in wavevector space. To quantify the anisotropic spectral distribution, we compute the two-dimensional power spectral density (2D-PSD) by averaging the spectral energy over concentric rings in the $k_x$--$k_y$ plane at fixed $k_\perp$. Specifically, we define the axisymmetric 3D spectrum of velocity field as:

\begin{equation}
\mathrm{PSD}(k_\perp, k_\parallel) = \sum_{k_\perp = \sqrt{k_x^2 + k_y^2}} \left| \tilde{v}(k_x, k_y, k_z) \right|^2 
\end{equation}

where $\tilde{v}(k_x, k_y, k_z)$ denotes the velocity fluctuation of the fast-mode, obtained by applying 3D Fourier transform to the velocity field $\mathbf{v}(x,y,z)$ followed by mode decomposition, $k_\perp = \sqrt{k_x^2 + k_y^2}$ is the perpendicular wavenumber, and $k_\parallel = k_z$ denotes the parallel wavenumber.

The one-dimensional spectrum can be obtained by integrating the 2D-PSD over one of the wavevector directions. For example, the perpendicular 1D spectrum is computed by integrating over the parallel direction:

\begin{equation}\label{eq_1D_specturm_perp}
\mathrm{PSD}_{1\mathrm{D}}(k_\perp) = \sum_{k_\parallel} \mathrm{PSD}(k_\perp, k_\parallel)
\end{equation}

Similarly, the parallel 1D spectrum is given by:

\begin{equation}\label{eq_1D_specturm_para}
\mathrm{PSD}_{1\mathrm{D}}(k_\parallel) = \sum_{k_\perp} \mathrm{PSD}(k_\perp, k_\parallel)
\end{equation}

For the 2D PIC simulation, we compute the two-dimensional spectrum of the velocity field within the $\mathbf{k}$--$\mathbf{B_0}$ plane, which corresponds to the compressible component:

\begin{equation}
\mathrm{PSD}(k_\perp, k_\parallel) = \mathrm{PSD}(k_y, k_x) = \left| \tilde{v}(k_x, k_y) \right|^2 
\end{equation}

where $\tilde{v}(k_x, k_y)$ denotes the velocity fluctuation of the fast-mode, obtained by applying 2D Fourier transform to the velocity field $\mathbf{v}(x,y)$ followed by mode decomposition. The 1D specturm is further calculated from Equations \ref{eq_1D_specturm_perp} and \ref{eq_1D_specturm_para}.

\section{Wave-mode decomposition}\label{appendix_decomp}
We adopt a local $\mathbf{k}$–$\mathbf{B}_0$ coordinate system, where: the parallel direction is defined along the background magnetic field $\mathbf{B}_0$: $\hat{\mathbf{e}}_\parallel = \hat{\mathbf{B}}_0$. The first perpendicular direction is given by: $\hat{\mathbf{e}}_{\perp1} = \hat{\mathbf{k}} \times \hat{\mathbf{B}}_0$. The second perpendicular direction is: $\hat{\mathbf{e}}_{\perp2} = \hat{\mathbf{B}}_0 \times (\hat{\mathbf{k}} \times \hat{\mathbf{B}}_0)$.

In this basis, the fast and slow magnetosonic modes have fluctuations confined to the $\mathbf{k}$–$\mathbf{B}_0$ plane, so their component along $\hat{\mathbf{e}}_{\perp1}$ vanishes. The Alfv\'en mode has fluctuations polarized in the $\hat{\mathbf{e}}_{\perp1}$ direction, which lies outside the $\mathbf{k}$–$\mathbf{B}_0$ plane.

The displacement vector for fast (+) and slow (-) modes is given by \citep{cho2002compressible,cho2003compressible,zhao2024small}:
\begin{equation}
    \boldsymbol{\xi}_\pm \propto \left( -1 + \alpha \pm \sqrt{A} \right) k_\parallel \hat{\mathbf{e}}_\parallel + \left( 1 + \alpha \pm \sqrt{A} \right) k_{\perp2} \hat{\mathbf{e}}_{\perp2}.
\end{equation}

The corresponding unit displacement vector is:
\begin{equation}
    \hat{\boldsymbol{\xi}}_\pm = \frac{\boldsymbol{\xi}_\pm}{|\boldsymbol{\xi}_\pm|}.
\end{equation}

The parameter $ A $ is defined as:
\begin{equation}
    A = (1 + \alpha)^2 - 4\alpha \cos^2 \theta, \qquad \alpha = \frac{c_s^2}{v_A^2},
\end{equation}
where $ v_A $ is the Alfv\'en speed, $ c_s $ is the sound speed, and $ \theta $ is the angle between the wavevector $ \mathbf{k} $ and $ \mathbf{B}_0 $.

The velocity perturbation amplitude along the direction of $ \mathbf{k} $ is then expressed as:
\begin{equation}
    \delta V_{\mathbf{k},\pm} \propto 
    \sqrt{D_{V_\parallel}(\mathbf{k})} \cos \zeta_{\hat{\mathbf{e}}_\parallel \hat{\boldsymbol{\xi}}_\pm}
    \pm \sqrt{D_{V_{\perp2}}(\mathbf{k})} \sin \zeta_{\hat{\mathbf{e}}_\parallel \hat{\boldsymbol{\xi}}_\pm},
\end{equation}
where $ \zeta_{\hat{\mathbf{e}}_\parallel \hat{\boldsymbol{\xi}}_\pm} $ is the angle between the unit vectors $ \hat{\mathbf{e}}_\parallel $ and $ \hat{\boldsymbol{\xi}}_\pm $, and $ D_V $ denotes the velocity variance in Fourier space along the corresponding direction.









\section{Theoretical turbulence spectrum equation}
\label{appendix_theory}

Starting from the turbulence energy equation \citep{petrosian2006damping}:

\begin{equation}
\frac{\partial W}{\partial t} = \frac{\partial}{\partial k_i} \left(D_{ij}\frac{\partial W}{\partial k_j} \right) - \Gamma(\textbf{k}) W - \frac{W}{T^W_{\mathrm{esc}}(\textbf{k})} + \dot{Q}^W.
\end{equation}

In a stationary state, $\partial W / \partial t = 0$, omitting the last two source and leakage terms on the right-hand side:

\begin{equation}\label{eq3}
\frac{\partial}{\partial k_i} \left(D_{ij}\frac{\partial W}{\partial k_j} \right) - \Gamma(\textbf{k}) W = 0.
\end{equation}

For fast mode turbulence, the cascade is assumed to be along the radial direction \citep{cho2002compressible,suzuki2007cascading,zhao2022multispacecraft}. Therefore, after transforming Equation \ref{eq3} into polar coordinate, we omit the term representing angular cascade and only keep the radial cascade item:

\begin{equation}\label{eq5}
\frac{1}{k} \frac{\partial}{\partial k} \left( k D_{kk}(k, \theta) \frac{\partial W(k, \theta)}{\partial k} \right)
=
\Gamma(k, \theta) W(k, \theta).
\end{equation}

For the left-hand side of Equation \ref{eq5}, $D_{kk} = k^2\tau_{\mathrm{cas}}^{-1}$.

Based on \citep{cho2003compressible,yan2004cosmic},  the cascade rate for fast mode turbulence is

\begin{equation}\label{eq77}
\tau_{\mathrm{cas}}^{-1} \sim \left(\frac{k}{L_0}\right)^{1/2}\frac{\delta V^2}{V_{\mathrm{ph}}},
\end{equation}

In \citet{galtier2023fast}, anisotropic cascading along different $\mathbf{k}$ directions is considered, owing to the weaker compressibility of fast modes propagating quasi-parallel to the background field. Equation (5.43) in \citet{galtier2023fast} gives the theoretical spectrum,  
$E = \sqrt{\frac{b_0\epsilon}{K_{\theta}}}\,C_K\,k^{-3/2},$
with the angular dependence coefficient is  
$\sqrt{K_\theta} = \frac{1+2\sin^2\theta}{\sin\theta}$.

From dimensional analysis, using $\delta v^2/\tau_{cas} = \mathrm{const}$, one obtains  
$Ek/\tau_{cas} = \sqrt{\frac{b_0\epsilon}{K_{\theta}}}\,C_K\,k^{-3/2}\cdot k/\tau_{cas} = \mathrm{const}$,
which leads to $
\tau_{cas}^{-1} \sim \sqrt{K_{\theta}}\,k^{1/2}$.

Combining with Equation \ref{eq77}, the cascade rate for fast-mode turbulence can be modified as  
\begin{equation}\label{eq7}
\tau_{\mathrm{cas}}^{-1} \sim C\left(\frac{1 + 2\sin^2\theta}{\sin\theta}\right) \left(\frac{k}{L_0}\right)^{1/2}\frac{\delta V^2}{V_{\mathrm{ph}}}.
\end{equation}
where C is a constant parameter to be determined. When $\theta=0$, we set the cascade rate $\tau_{cas}^{-1}=0$.

Thus,

\begin{equation}\label{eq8}
D_{kk}(k, \theta) = C\left(\frac{1 + 2\sin^2\theta}{\sin\theta}\right)(\frac{1}{L_0})^{1/2}\frac{\delta V^2}{V_{\mathrm{ph}}}k^{5/2}\\
\equiv C_0(\theta) k^{5/2}.
\end{equation}

For the right-hand side of Equation \ref{eq5}, we take the damping rate \citep{yan2004cosmic,ginzburg1962propagation}, which is valid at $\beta < 1$, as :

\begin{equation}\label{eq10}
\Gamma = \frac{\sqrt{\pi\beta}}{4}\frac{\sin^2\theta}{\cos\theta} \left[\sqrt{\frac{m_e}{m_p}} \exp\left(-\frac{m_e}{m_p\beta\cos^2\theta}\right) + 5 \exp\left(-\frac{1}{\beta\cos^2\theta}\right)\right] V_{\textrm{ph}} k \equiv C_1(\theta) k.
\end{equation}

Substituting Equations \ref{eq8} and \ref{eq10} into Equation \ref{eq5}, we obtain:

\begin{equation}\label{eq12spectrum}
\frac{\partial}{\partial k} \left(C_0(\theta) k^{7/2} \frac{\partial W}{\partial k} \right)
=
C_1(\theta) k^2 W.
\end{equation}

We propose a spectrum of the form:
\begin{equation}
W(k,\theta) \approx k^{-\alpha} e^{-S(k,\theta)},
\end{equation}
where $\alpha$ captures the underlying power-law behavior and $S(k,\theta)$ captures additional damping effects.

The left-hand side of Equation \ref{eq12spectrum} becomes:
\begin{equation}
\frac{d}{dk} \left( C_0(\theta) k^{7/2} \frac{dW}{dk} \right) = \frac{7}{2}C_0(\theta) k^{5/2}\frac{dW}{dk} + C_0(\theta) k^{7/2} \frac{d^2W}{dk^2},
\end{equation}
and the right-hand side of Equation \ref{eq12spectrum} is:
\begin{equation}
C_1(\theta) k^2 W = C_1 k^2 \cdot k^{-\alpha} e^{-S(k)}.
\end{equation}

Divide both sides by $k^{-\alpha} e^{-S(k)}$, then simplify to get:
\begin{equation}\label{eq27}
k^{3/2}(\alpha^2 - \frac{5}{2}\alpha) + k^{5/2}(2\alpha - \frac{7}{2}) \frac{dS}{dk}  + k^{7/2} \left( \left(\frac{dS}{dk}\right)^2 - \frac{d^2S}{dk^2} \right) = \frac{C_1}{C_0} k^2.
\end{equation}

Assuming that the second derivative of $ S(k) $ is small and can be neglected, we obtain a solution to the simplified equation for $ \alpha = 5/2 $ in 2D situation as:
\begin{equation}
u(k,\theta)\equiv \frac{dS}{dk}=\frac{\sqrt{4\frac{C_1}{C_0}k^{1/2}+\frac{9}{4}}-\frac{3}{2}}{2k}.
\end{equation}

However, in this case, we find that $ u^2 \ll \frac{du}{dk} $ ($i.e., (\frac{dS}{dk})^2 \ll \frac{d^2S}{dk^2} $) for $ k \ll 1 $, indicating that the second derivative of $ S(k) $ should be retained, while the term $ \left(\frac{dS}{dk}\right)^2 $ can instead be neglected.

Neglecting $ \left(\frac{dS}{dk}\right)^2 $, Equation \ref{eq27} reduces to:
\begin{equation}
k\frac{du}{dk} - \frac{3}{2}u + \frac{C_1}{C_0}k^{-1/2} = 0.
\end{equation}

Solving this equation for $ u $, we obtain:
\begin{equation}
u = \frac{C_1}{2C_0}k^{-1/2} + C_2k^{3/2},
\end{equation}
where the second term on the right-hand side is independent of $ \theta $. Therefore, we just set the integration constant $ C_2 = 0 $.

Integrating $u$ to obtain $S(k,\theta)=\int udk$, the spectrum in the 2D situation becomes:
\begin{equation}
W(k,\theta) \approx k^{-5/2} e^{-\frac{C_1}{C_0}k^{1/2}} = k^{-5/2} e^{-\frac{\Gamma}{\tau_{cas}^{-1}}}.
\end{equation}

This spectrum recovers $W(k,\theta)\approx k^{-5/2}$  in the no-damping case $\Gamma \to 0$, and to the common $W(k)\propto k^{-3/2}$ after integration over $\theta$.


\end{document}